# Ab-initio calculations of magnetic and optical properties of 3x3 supercell of TiX$_2$ (X = S, Se and Te) compounds under the effect of externally applied electric field and strain

*Aditya Dey*

Department of Physics, Indian Institute of Technology Patna, Bihta Campus, India

*The magnetic and optical properties of titanium dichalcogenide compounds, TiX$_2$ (X = S, Se and Te) have been calculated by first principles calculations using density functional theory (DFT) as implemented in SIESTA code. A 3x3 supercell of the compounds is taken in this study to be able to tune the properties obtained from the unit cell of these compounds. With magnetic states (ferromagnetic nature) as the attained ground state, spin polarized calculations are performed to obtain the mentioned properties. Further, with spin polarization into effect, external electric field (along z-direction) and biaxial strain (along x and y-directions) is employed to study their effect on these properties. The effect of biaxial strain on the geometry of the compounds is also studied. It is observed that the pristine supercell of the compounds possess a good amount of magnetic moment and this can be modulated using the applied field and strain. For the optical properties, polarized light along the z-direction (c axis) is used. These properties include the calculation of real & imaginary parts of dielectric function, absorption coefficient, reflectance, optical conductivity and refractive index in 0-25 eV energy range. Various modulations of these properties are observed including the blue-shifts and red-shifts of energies with highest peaks in the visible region and also shifting of energies to other regions of the electromagnetic spectrum. Hence, with the tunable magnetic and diverse optical properties, the compounds can be useful in the field of spintronics and in making various optical devices.*

## 1. Introduction

In recent studies of material science and related areas over a span of few years, two-dimensional materials are in the limelight due to their variety of properties and remarkable applications in numerous fields. Of these materials, potential research is being carried out on layered transition metal dichalcogenides (TDMC) because of its anisotropic physical properties and strong covalent bound layers as well as their different technological applications [1-6].The titanium dichalcogenide compounds, TiX$_2$ (X = S, Se and Te) are one of the types of TDMC which has attracted a great amount of attention due to its wide range of applications [7-13]. These compounds comprise of a covalently bonded Ti and X atoms in two-dimensional hexagonal planes. The planes are stacked together by chalcogen-chalcogen van der Waals interactions [14-16]. These compounds have highly anisotropic physical properties. Therefore, significant changes in their properties can be made by foreign atoms or external factors.

TiX$_2$ crystallizes in 1*T*-type structure and with a space group *P*-3*m*1 at room temperature and pressure [17]. In earlier research works, these compounds have been studied both experimentally and theoretically including the study on their various properties. With an emphasis on the theoretical studies of these compounds, many researchers have dug into several properties like electronic, mechanical, magnetic, optical properties and so on [18-21]. It is observed that all of these compounds possess a non-magnetic behaviour with 0.00 µ$_B$ magnetic moment. For the optical properties, transmission spectra of TiS$_2$ and TiSe$_2$ are measured at liquid-helium temperature is measured with $E \perp c$ in the energy range 0.5–4 eV [22]. Also the reflectivity of single crystals of TiS2, TiSe2 and TiTe2 is measured with the electric field perpendicular to the hexagonal *c* axis in the energy range 1–12 eV [23]. In addition to this, absorption spectra and reflectivity in other energy ranges has also been reported [24-25]. Mentioning about the electronic properties, the previous reports show that there are many conflicts on their study as some works have reported that they behave as a semiconductor and at the same time some classify it as a semi-metal [26-28]. Hence, it remains as a matter of dispute as the property seems to change due to some marginal factors depending on different parameters. Also, some of reports show that people have studied the effect of strain or pressure on the band structures of these compounds in order to modulate the electronic properties [29-31].

Going through the available literature based on the study of these compounds, it can be found out that there is very scarce work done on the study of possible changes in their optical and magnetic properties which can be caused by external factors. So, in this article, a study has been performed to investigate the possible modulation and changes in the optical properties and magnetic moment of $TiX_2$ compounds via the effect of external electric field and strain, which is done theoretically using first principles calculation by density functional theory. In view of the optical properties, various optical characteristics of the compounds like absorption coefficient, reflectance, real and imaginary parts of dielectric function, refractive index and optical conductivity has been included in the study.

## 2. Computational details

The first principle calculations are performed using density functional theory (DFT) as implemented in the ab-initio package SIESTA [32-36]. The approach is based on an iterative solution of the Kohn–Sham equation. The generalized-gradient approximation (GGA) with the Perdew-Burke-Ernzerhof (PBE) form [37] is chosen for the exchange-correlation functional. A real space mesh cutoff of 300 Ry has been used for representing the charge density. Conjugate-gradient (CG) method [38] has been used to optimize the structures. Normconserving pseudopotentials in the fully nonlocal Kleinman−Bylander form have been considered for all the atoms [39]. A double-ζ polarized (DZP) basis set is used. Systems are considered to be relaxed only when the forces acting on all the atoms are less than 0.01 eV Å$^{-1}$. A 3x3 supercell is taken for all the three compounds which are periodic along x and y directions (*a* & *b* axis) and vacuum space of atleast 20 Å is created along z direction to avoid any spurious interactions between the layers. The convergence for energy was chosen as $10^{-4}$ eV between two steps. Firstly, all the supercell structures of the compounds were optimized in both magnetic and non-magnetic states. Finding magnetic states as the ground state spin polarized self-consistent calculations were performed to employ the electric field along *c* axis and magnetic moment is obtained. Again using spin polarization, strain along *a* and *b* axis is applied and the structures are optimized followed by calculating the moment. Further, after applying external field and strain, optical properties were calculated for out of plane polarized light (E||c) for an energy range of 0-25 eV and the direction of optical vector being (001). Optical calculations were carried out using 40 X 40 X 2 optical mesh and optical broadening used is 0.1 eV. While performing the calculations of the mentioned properties under the effect of external factors and optimization of the structures as well, a 15 X 15 X 1 k-point grid has been considered within the Monkhorst−Pack scheme [40] for each and every calculation.

## 3. Results and Discussions

### 3.1 Structural parameters

As stated previously, the main aim here is to study the effect of externally applied electric field and strain on the optical properties and magnetic moment of the $TiX_2$ compounds. For that, a 3 X 3 supercell of each of these compounds has been taken. The supercell is periodic only in two directions (*a* and *b* axis or x & y directions) and non-periodic along the *c*-axis. The structures have been optimized in non-magnetic states followed by optimization including spin polarization. Table 1 shows the comparison between total energies of both magnetic (M) and non-magnetic (NM) states.

From the table, it can be seen that the energy difference between the M and NM states ($E_M - E_{NM}$) is negative for all the structures of $TiX_2$, indicating that the magnetic state is more stabilized over the non-magnetic state. Also as the stabilization energy is much higher than the room temperature energy (0.025 eV), M state seems to be robust in room temperature.

| Compounds | $E_{NM}$ (eV) | $E_M$ (eV) | $E_M - E_{NM}$ (eV) | a(Å) | b(Å) | c(Å) | α(deg) | β(deg) | γ(deg) |
|---|---|---|---|---|---|---|---|---|---|
| $TiS_2$ | -6732.917 | -6735.880 | -2.963 | 9.23 | 9.22 | 20.59 | 89.74 | 89.96 | 120.00 |
| $TiSe_2$ | -6042.621 | -6046.684 | -4.063 | 9.27 | 9.28 | 21.05 | 89.99 | 90.05 | 120.04 |
| $TiTe_2$ | -5042.268 | -5047.297 | -5.029 | 10.03 | 10.03 | 21.77 | 90.03 | 90.04 | 120.01 |

Table -1 Comparison of total energies of M and NM states and spin-polarized optimized structural parameters of $TiX_2$ compounds

A schematic model of the spin-polarized (M) optimized structure of these compounds is shown in Figure 1, which shows both the top view and side view of the supercell. It must be noted that the shown structure is same for all the compounds with slight changes in the lattice parameters, as recorded in Table 1. In one of the previously reported works on $WSe_2$ [41], it is observed that the spin-orbit interaction has almost no effect on the optical properties. So, in accordance to this report, in this study also, calculations have been performed including the spin orbit coupling (SOC), and similarly it is observed that there is very minor change in the results and so it is not reported in this article. While SOC is an essential factor for spin-polarized states in heavy metal compounds like of titanium, but the spin polarization emerging due to the magnetic moment of the atoms and the effect of SOC due to heavy metal (Ti in this case) plays a very minor role in the net magnetic moment of the compounds.

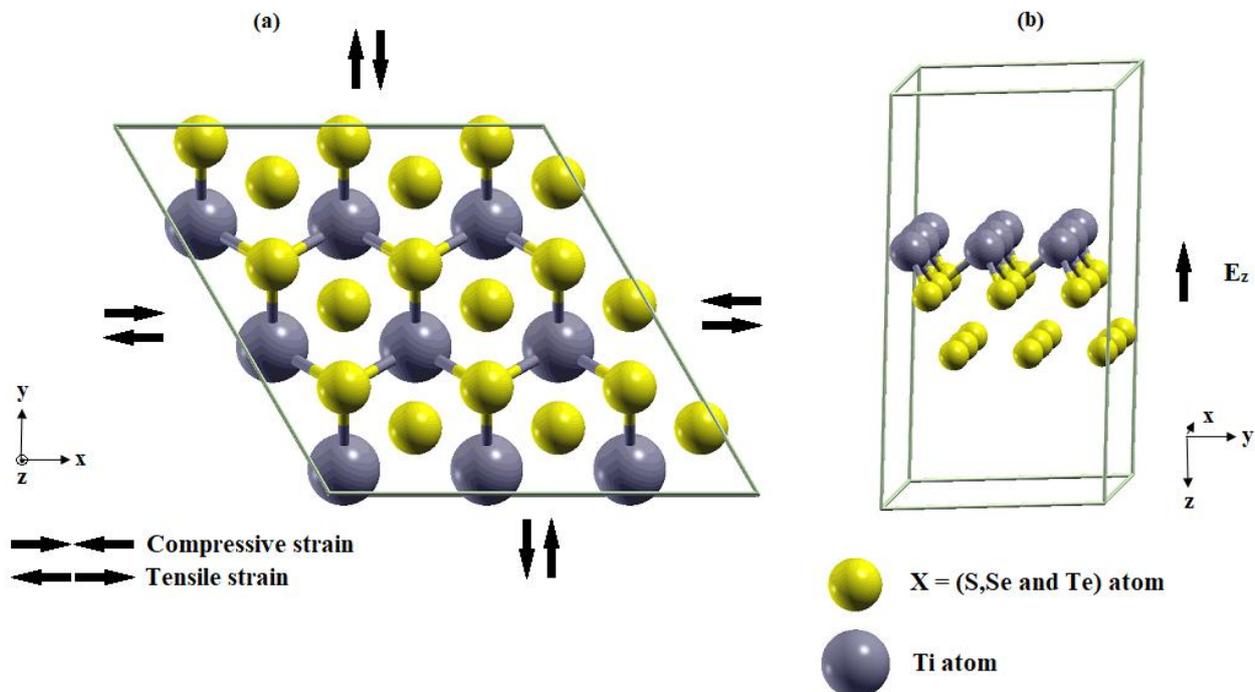

Figure 1 Ball and stick schematic diagram of $TiX_2$ compounds with (a) top view (b) side view including the directions of applied electric field and strain

The schematic above also shows the direction in which the externally applied electric field and biaxial strain is applied. Electric field is applied in the compounds in out of the plane or z direction (along *c* axis) which is denoted as $E_z$. The magnitude of field strength is varied from 2.5 V/nm to 10

V/nm. The optical properties as well as the magnetic moment is calculated and studied with different magnitudes of $E_z$. Self-consistent spin polarized calculation with particular magnitude of the field strength is performed to study the effect of $E_z$ on the mentioned properties.

Strain is applied simultaneously along both *a* and *b* axis (x and y-directions) as shown in Figure 1. The biaxial strain applied is both tensile and compressive, denoted by plus (+) and minus (-) sign respectively. The amount of applied strain is ±2%, ±5% and ±10%. After freezing the strained lattices, the structures are optimized with spin polarization into effect and the geometric variations of the optimized structures are noted in Table 2. It can be seen that for all the compounds, Ti-X distance increases under tensile strain and decreases under compression. Whereas, for interlayer distance between S atoms, the previous trend is just reversed. This is in good agreement with a previous study done for strain in $TiS_2$ [42]. With increase in the atomic number of chalcogen atom, both $d_{Ti-S}$ and $h_{S-S}$ increases.

|  | $TiS_2$ | | $TiSe_2$ | | $TiTe_2$ | |
| --- | --- | --- | --- | --- | --- | --- |
| **Strain (%)** | $d_{Ti-S}$ | $h_{S-S}$ | $d_{Ti-Se}$ | $h_{S-S}$ | $d_{Ti-Te}$ | $h_{S-S}$ |
| **Pristine 3X3 supercell** | 2.35 | 2.96 | 2.52 | 3.31 | 2.77 | 3.70 |
| **+ 2%** | 2.38 | 2.87 | 2.56 | 3.20 | 2.84 | 3.58 |
| **+ 5%** | 2.43 | 2.79 | 2.60 | 3.13 | 2.91 | 3.44 |
| **+ 10%** | 2.49 | 2.66 | 2.68 | 3.02 | 3.01 | 3.35 |
| **- 2%** | 2.31 | 3.04 | 2.41 | 3.43 | 2.72 | 3.77 |
| **- 5%** | 2.28 | 3.13 | 2.32 | 3.56 | 2.66 | 3.86 |
| **- 10%** | 2.25 | 3.21 | 2.19 | 3.68 | 2.60 | 3.98 |

Table 2 – Variations in geometry of the optimized structures under strain. $d_{Ti-X}$ denotes bond length between nearest Ti and X atom and $h_{S-S}$ denotes interlayer height between upper and lower S atomic planes

### 3.2 Magnetic properties

As can be found from the previous works, there is no magnetic moment present in the unit cell structure of the $TiX_2$ compounds. However, from this study it can be seen that a moment can be induced in these compounds by creating their supercell (refer Table 3 or Table 4), which is obtained because the structures show stabilization in the magnetic state (ferromagnetic arrangement). As can be seen from the tables, the obtained magnetic moment for pristine 3 X 3 supercell is 1.40 for $TiS_2$, 1.91 for $TiSe_2$ and 2.13 $\mu_B$ for $TiTe_2$, which is a considerable amount of moment. These moments are calculated per unit cell of the compounds. It is deduced by the expression, $\mu_{TiX2} = \mu_{Ti} - 2(\mu_X)$, where $\mu_{Ti}$ denotes the magnetic moment of Ti atom and $\mu_X$ denote that of chalcogen atoms (S, Se & Te atoms). $\mu_{TiX2}$ denotes the total moment per unit cell of the compounds.

Also, this magnetic moment can be varied depending on the different magnitude of electric field and strain. Table 3 and Table 4 show the total magnetic moment and the local magnetic moment (per unit cell) as well, which is the individual contribution of Titanium and the chalcogen atom to the total moment, for all the compounds of $TiS_2$, $TiSe_2$ and $TiTe_2$ with the applied field and strain respectively. It can be seen that the contribution to the total moment is majorly by the Ti atom, while the magnetic moment due to the chalcogen atoms is negligible. The electronic configuration of Ti atom is [Ar] $3d^24s^2$ and that of S is [Ne] $3s^23p^4$, Se is [Ar] $3d^{10}4s^24p^4$ and Te is [Kr] $4d^{10}5s^25p^4$. Hence, the magnetic contributions of Ti is mainly due to 3d orbital electrons and the somewhat negligible amount of moment due to S, Se and Te atoms is due to p orbital electrons. The reason for this is that for s- and p- valence electrons, the bonding-antibonding splitting is large which results in translating to a large band width due to strong orbital overlap in a solid. Formation of magnetic states is then unfavourable because large band width is equivalent to small density of states. In contrast, the spatial extent of d and f orbitals is much smaller and also these orbitals only are enough localized and narrow as compared to other orbitals to give a high value to the exchange integrals between the spins, due to which they produce considerable amount of magnetic moment. Also, it is interesting to note that the magnetic moment contributed by the chalcogen atom is negative, which indicates that the chalcogen magnetic moment coming from the S 3p, Se 4p and Te 5p states is anti-parallel to the magnetic moment contributed by the Ti 3d states. Next, the effect of $E_z$ and biaxial strain applied on these compounds is shown. It is observed that among the compounds, TiTe$_2$ has highest moment in pristine supercell as well as with the applied $E_z$ and strain and TiS$_2$ has the lowest values.

**Effect of electric field**

Table 3 shows the variation of magnetic moment with each particular magnitude of applied field strength, which is applied in out of the plane direction. The attained magnetic behaviour for these compounds is of ferromagnetic nature.

| | TiS$_2$ | | | TiSe$_2$ | | | TiTe$_2$ | | |
|---|---|---|---|---|---|---|---|---|---|
| **Electric field ($E_z$)** | $\mu_{Ti}$ | $\mu_S$ | $\mu_{TiS2}$ | $\mu_{Ti}$ | $\mu_{Se}$ | $\mu_{TiSe2}$ | $\mu_{Ti}$ | $\mu_{Te}$ | $\mu_{TiTe2}$ |
| **Pristine 3X3 supercell** | 1.80 | -0.20 | **1.40** | 2.29 | -0.19 | **1.91** | 2.32 | -0.09 | **2.13** |
| **2.5 V/nm** | 1.88 | -0.11 | **1.66** | 2.39 | -0.16 | **2.07** | 2.57 | -0.05 | **2.47** |
| **5 V/nm** | 1.85 | -0.10 | **1.65** | 2.36 | -0.15 | **2.06** | 2.51 | -0.03 | **2.45** |
| **7.5 V/nm** | 1.83 | -0.11 | **1.61** | 2.35 | -0.15 | **2.05** | 2.50 | -0.03 | **2.44** |
| **10 V/nm** | 1.84 | -0.13 | **1.58** | 2.33 | -0.15 | **2.03** | 2.47 | -0.02 | **2.43** |

Table -3 Magnetic moments of TiX$_2$ compounds with and without (pristine 3X3 supercell) $E_z$. (All moments are in Bohr magneton, ($\mu_B$)).

Now, by application of electric field, it is observed that there is some amount of increase in the magnetic moment for all the compounds, with the highest increase in TiTe$_2$ with a total moment 2.47 $\mu_B$, when the $E_z$ magnitude is 2.5 V/nm and the same for TiS$_2$ is 1.66 and 2.07 $\mu_B$ for TiSe$_2$. Again, when the magnitude of field strength $E_z$ is increased, a trend of gradual decrease in the values of moment is recorded. It reduces to 1.58 $\mu_B$ for TiS$_2$, 2.03 $\mu_B$ for TiSe$_2$ and 2.43 $\mu_B$ for TiTe$_2$ at 10 V/nm. So, it is found that employing $E_z$, it is possible to increase the magnetic moment and even though the moment is decreasing with increasing $E_z$, the values are still greater than that of pristine TiX$_2$ supercell. Although the decrease is marginal, the trend is what must be noted. It is well known that by applying electric field to a compound or an element, there is subtle movement of the electron charge density. Hence it is possible that the electrons jump from one orbital to the other. Also, the partially filled d orbitals tend to show better ferromagnetism. Hence, in this case, in Ti atom, whose magnetism is generated by 3d states, the electric field causes movement of electron from s to d orbital, enabling it to show more magnetic moment and so we observe increase in local magnetic moment of Ti ($\mu_{Ti}$). Along with Ti, though marginal, there is also shift of electrons in chalcogen atoms ($\mu_X$). With increase in $E_z$, due to more variation in charge density in both atoms, net moment is getting reduced.

**Effect of strain**

Table 4 shows the modulation of magnetic moment of the TiX$_2$ compounds with the percentage of biaxial strain applied along *a* and *b* axis (+ indicates tensile and – indicates compressive strain). As stated earlier, the magnetic nature obtained is ferromagnetic.

| Strain (%) | TiS$_2$ | | | TiSe$_2$ | | | TiTe$_2$ | | |
|---|---|---|---|---|---|---|---|---|---|
| | $\mu_{Ti}$ | $\mu_S$ | $\mu_{TiS2}$ | $\mu_{Ti}$ | $\mu_{Se}$ | $\mu_{TiSe2}$ | $\mu_{Ti}$ | $\mu_{Te}$ | $\mu_{TiTe2}$ |
| Pristine 3X3 supercell | 1.80 | -0.20 | **1.40** | 2.29 | -0.38 | **1.91** | 2.32 | -0.09 | **2.23** |
| + 2% | 1.35 | -0.13 | **1.09** | 1.42 | -0.31 | **1.11** | 1.41 | -0.31 | **1.10** |
| + 5% | 0.62 | -0.12 | **0.38** | 1.33 | -0.34 | **0.99** | 1.22 | -0.26 | **0.96** |
| + 10% | 0.61 | -0.14 | **0.33** | 1.02 | -0.38 | **0.64** | 1.12 | -0.35 | **0.77** |
| - 2% | 1.81 | -0.18 | **1.45** | 1.97 | -0.04 | **1.93** | 2.45 | -0.04 | **2.41** |
| - 5% | 1.87 | -0.16 | **1.55** | 2.32 | -0.13 | **2.19** | 2.56 | -0.09 | **2.47** |
| - 10% | 1.95 | -0.14 | **1.67** | 2.44 | -0.08 | **2.36** | 2.63 | -0.10 | **2.53** |

Table – 4 Magnetic moments of TiX$_2$ compounds with and without (pristine 3X3 supercell) strain. (All moments are in Bohr magneton, ($\mu_B$)).

For TiS$_2$, total moment decreases from 1.40 to 1.09 $\mu_B$ at +2% strain and tends to become feeble at +5% and +10% strain with values 0.38 and 0.33 $\mu_B$ respectively. Again it increases to 1.45 $\mu_B$ at -2% strain and climbs up to 1.55 and 1.67 $\mu_B$ at -5% & -10% respectively. Similarly, for TiSe$_2$, from 1.91 $\mu_B$, it decreases to 1.11, 0.99 & 0.64 $\mu_B$ for +2%, +5% and +10% strain and increases to 1.93, 2.19 & 2.36 $\mu_B$ for -2%, -5% and -10% strain respectively. Again for TiTe$_2$, it decreases from 2.23 $\mu_B$ to 1.10, 0.96 & 0.77 $\mu_B$ for +2%, +5% and +10% strain & increases to 2.41, 2.47 & 2.53 $\mu_B$ for -2%, -5% and -10% strain respectively. So, it is seen that when tensile strain is applied, there is sudden decrease in the moment as compared to the unstrained structures, of both the local and total magnetic moment for all the compounds. Gradually, by increasing the percentage of strain, the total magnetic moment values become almost negligible. On the other hand, by employing compressive strain, there is a gradual increase in the values, and both the total and local magnetic moments increase with increasing the compressive strain. This is because due to tensile strain, the inter-atomic distance is increased and thus there is decrease in the exchange correlation or small overlap between the atomic orbitals. As a result, the ferromagnetic response gets suppressed, thus reducing the magnetic moment. With a similar explanation, as the inter-atomic distance decreases due to compressive strain, the magnetic moment gets enhanced. Thus, magnetic moment can be tuned by applying strain.

### 3.3 Optical properties

As stated earlier, the optical properties of all the compounds have been calculated for polarized light in out of the plane direction of the optical vector (001), that is, E ∥ $c$ axis in the energy range 0-25 eV. In view of this, different properties that have been investigated are absorption coefficient (α), reflectance (R), imaginary ($\varepsilon_2$) and real ($\varepsilon_1$) parts of dielectric constant, refractive index (n) and optical conductivity (σ). The complex dielectric function ε (ω), can be expressed as ε (ω) = $\varepsilon_1$ (ω) + $i\varepsilon_2$ (ω) and optical conductivity σ (ω) is calculated using ε (ω) = $\varepsilon_0$ + $i\sigma$ (ω) / ω, where $\varepsilon_0$ is the vacuum permittivity. The deduced properties are shown for TiS$_2$ only and the properties of the other two compounds are reported and compared further.

**(a) Real and imaginary parts of dielectric function**

The real and imaginary parts of the dielectric function are calculated for all the compounds. The pristine supercell of TiS$_2$ has maximum $\varepsilon_1$ value 7.21 at 1.52 eV and maximum $\varepsilon_2$ = 11.42 at 1.57 eV (Figure 2). The same for TiSe$_2$ $\varepsilon_1$ = 10.72 at 1.18 eV and $\varepsilon_2$ = 12.75 at 1.31 eV and for TiTe$_2$, $\varepsilon_1$ = 13.08 at 1.42 eV and $\varepsilon_1$ = 13.51 at 1.81 eV (Table 5). For $\varepsilon_1$, there are some dips in the amplitude in the range 1.6-2.2 eV, where it attains negative value indicating that there is plasmonic excitation in that range and most of the incident electromagnetic wave is reflected in this region. It can be observed that peaks are seen in reflectance spectrum in the places where $\varepsilon_1$ becomes negative. Also, $\varepsilon_1$ becomes steady with increase in energy showing that it does not interact with energy photons (Figure 2(a)). A similar observation is noted for the other two compounds also. For $TiS_2$, strong peaks of both $\varepsilon_1$ and $\varepsilon_2$ appear at near IR region (1.2 – 1.8 eV) whereas for $TiSe_2$ and $TiTe_2$, it appears at both near IR and visible region (1.6 – 3.2 eV), which shows a contrasting behaviour of dielectric function with regard to electromagnetic spectrum. For all the compounds, peaks for $\varepsilon_2$ are observed in ~1.5-2 eV range, which suggest that the compounds absorb electromagnetic waves in this range for this direction of polarization. At high energy (> 20 eV), $\varepsilon_2$ values vanishes, that tells there is no absorption of the compounds at very high frequency. The static dielectric constant is 5.13, 5.18 and 5.32 for $TiS_2$, $TiSe_2$ & $TiTe_2$ respectively which shows that the value marginally increases with increasing atomic number of the chalcogen atom.

**Effect of electric field –** By employing $E_z$, a similar trend of gradual increase or decrease of the $\varepsilon_1$ & $\varepsilon_2$ values is found in the compounds. For $TiS_2$, $\varepsilon_1$ & $\varepsilon_2$ increase to 8.16 & 14.18 at 1.38 and 1.43 eV respectively at 5 V/nm and then decrease further (Figure 2). In case of $TiSe_2$, a drastic jump in the values is seen where $\varepsilon_1$ & $\varepsilon_2$ increase to 16.47 at 1.06 eV and 20.52 at 1.12 eV for 10 V/nm. For $TiTe_2$, $\varepsilon_1$ & $\varepsilon_2$ decrease to 12.04 & 12.14 at 1.23 & 1.36 eV respectively at 10 V/nm. Also, there is no major shift in energies is observed with the application of electric field. The increase or decrease of values is related to the movement of electron density and electronic transitions caused by the effect of external field. Although, the changes in static dielectric constant is quite less for $TiS_2$ as compared to the other two compounds. For $TiSe_2$ & $TiTe_2$, the values tend to increase gradually with increasing $E_z$ (Table 5).

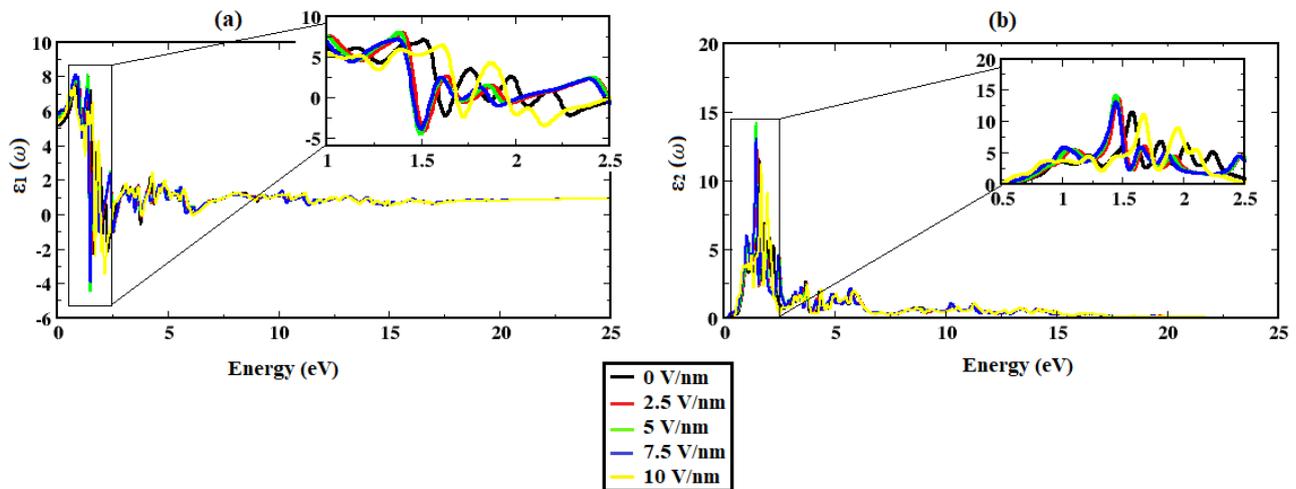

Figure 2 – (a) Real and (b) Imaginary values of dielectric function for $TiS_2$ under the effect of electric field. The peak regions are shown in the inset.

| | $TiS_2$ | | | $TiSe_2$ | | | $TiTe_2$ | | |
|---|---|---|---|---|---|---|---|---|---|
| $E_z$ (V/nm) & Strain (%) | Energy at max. peak (eV) ; $\varepsilon_1$ | Energy at max. peak (eV) ; $\varepsilon_2$ | Static dielectric const. | Energy at max. peak (eV) ; $\varepsilon_1$ | Energy at max. peak (eV) ; $\varepsilon_2$ | Static dielectric const. | Energy at max. peak (eV) ; $\varepsilon_1$ | Energy at max. peak (eV) ; $\varepsilon_2$ | Static dielectric const. |
| Pristine supercell | 1.52; **7.21** | 1.57; **11.42** | 5.13 | 1.18; **10.72** | 1.31; **12.75** | 5.18 | 1.42; **13.08** | 1.81; **13.51** | 5.32 |
| 2.5 V/nm | 1.41; **7.95** | 1.46; **13.76** | 5.52 | 1.15; **13.2** | 1.25; **16.45** | 5.57 | 1.34; **12.85** | 1.52; **13.09** | 5.70 |
| 5 V/nm | 1.38; **8.16** | 1.43; **14.18** | 5.64 | 1.12; **14.54** | 1.19; **17.90** | 5.78 | 1.30; **12.54** | 1.51; **12.77** | 5.89 |
| 7.5 V/nm | 1.37; **7.18** | 1.43; **13.06** | 5.81 | 1.11; **14.59** | 1.16; **19.23** | 6.01 | 1.27; **12.27** | 1.50; **12.05** | 6.06 |

| 10 V/nm | 1.61; **6.38** | 1.67; **11.06** | 5.42 | 1.06; **16.47** | 1.12; **20.52** | 6.19 | 1.23; **12.04** | 1.36; **12.14** | 6.32 |
| +2 % | 2.68; **2.73** | 2.80; **1.45** | 2.20 | 0.02; **4.12** | 3.15; **1.39** | 4.08 | 0.62; **3.14** | 4.09; **1.47** | 2.93 |
| +5 % | 2.21; **2.48** | 4.68; **1.30** | 2.08 | 2.02; **2.82** | 4.19; **1.41** | 2.32 | 0.34; **3.53** | 3.57; **1.42** | 3.31 |
| +10 % | 2.19; **2.46** | 4.98; **1.35** | 2.10 | 2.08; **2.68** | 10.28; **1.27** | 2.11 | 1.24; **3.36** | 3.20; **1.38** | 2.70 |
| -2 % | 0.11; **5.38** | 3.56; **1.68** | 5.06 | 0.50; **3.29** | 4.36; **1.58** | 3.02 | 0.34; **3.94** | 1.60; **1.42** | 3.91 |
| -5 % | 3.36; **2.92** | 3.74; **1.58** | 5.40 | 0.00; **8.83** | 0.42; **3.47** | 8.83 | 0.00; **8.52** | 0.47; **3.37** | 8.52 |
| -10 % | 1.26; **3.13** | 5.27; **1.51** | 2.73 | 0.00; **9.98** | 0.28; **3.65** | 9.98 | 0.02; **9.15** | 0.14; **3.44** | 9.15 |

Table -5 Values of static dielectric constant and real & imaginary values of dielectric function including energies with peak positions under the effect of $E_z$ and strain for all the compounds.

**Effect of strain** – With the different amount of applied biaxial strain, it is seen that there are changes in the values of dielectric function along with shift in energies which varies with the direction of strain and the compound also. For $TiS_2$, both the $\varepsilon_1$ & $\varepsilon_2$ values are decreased to a great extent for both directions of strain (Figure 3). The energies for $\varepsilon_2$ values in both the strain are blue shifted & for $\varepsilon_1$, blue-shifts are present in tensile strain and red shifts are seen in -2% & -10% strain. The static dielectric constant is reduced in tensile strain & for -10% strains. From Table 5, one can observe that similar to $TiS_2$, for $TiSe_2$ & $TiTe_2$ also there are mixed variations of energy shifts and $\varepsilon_1$ & $\varepsilon_2$ values, however it must be noted that both $\varepsilon_1$ & $\varepsilon_2$ values have decreased with respect to that in pristine supercell with the application of strain. For tensile strain in $TiSe_2$ & $TiTe_2$, static dielectric constant decreases alike $TiS_2$ and also for compressive strain of -2 % but the value increases for -5% & -10% strain.

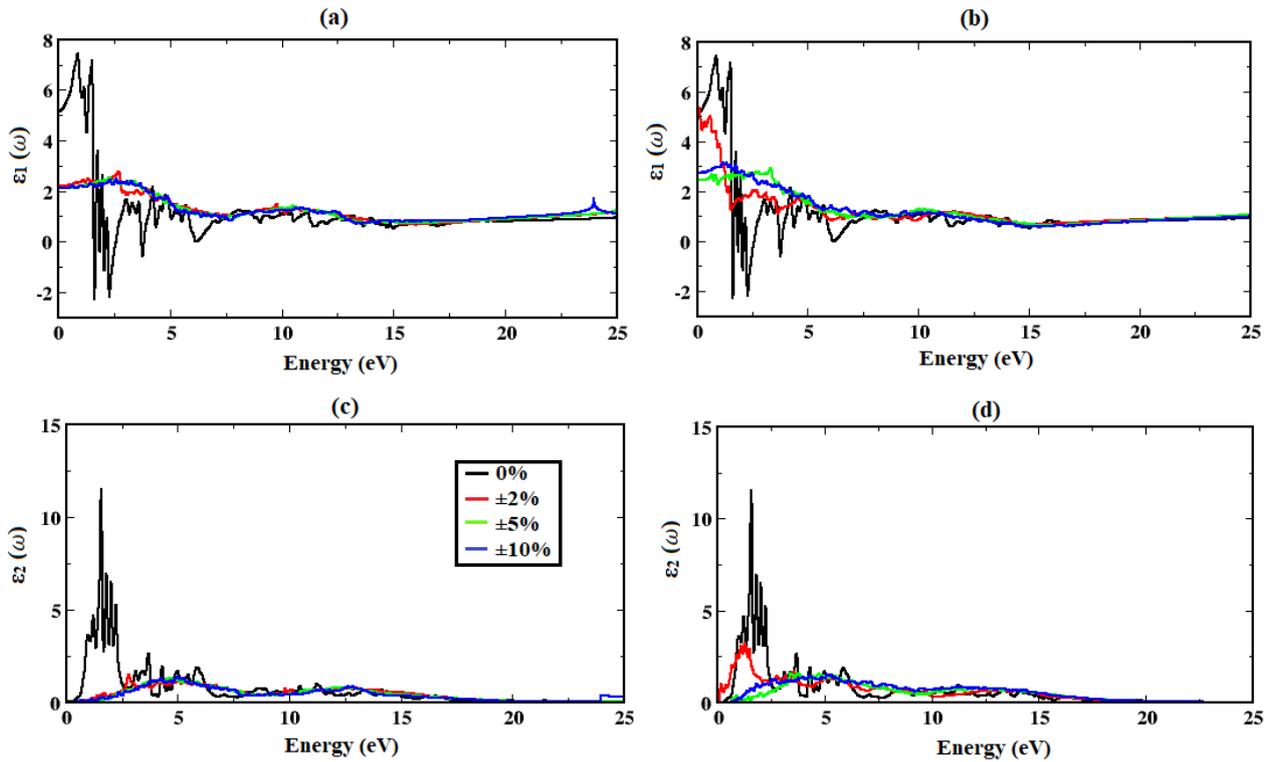

Figure – 3 Real and imaginary values of dielectric function for $TiS_2$ under (a, c) tensile and (b, d) compressive strain. The peak regions are shown in the inset.

**(b) Absorption coefficient and Reflectance**

The pristine supercell of $TiS_2$, that is, without any applied field or strain, shows that there is strong absorption at 6.06 eV and 11.37 eV, with the highest peak at 11.37 eV with $\alpha = 5.4 \times 10^5$ cm$^{-1}$ (Figure 4(a) or 5(a,b) & Table 6). Both the peaks show that they are in the Ultraviolet light region (3.9 – 14 eV photon energy), which means there are low electron losses and high absorbance in this region and can be attributed to the interband transitions, similar to as stated for $\varepsilon_2$. Similarly, the pristine $TiSe_2$ supercell show strong absorption at 10.34 and 13.24 eV, with maximum value of $\alpha = 10.4 \times 10^5$ cm$^{-1}$ at 13.24 eV and for $TiTe_2$ highest peaks are attained at 8.75 & 13.82 eV with maximum absorption of $\alpha = 6.5 \times 10^5$ cm$^{-1}$ at 13.82 eV. So, as similar to $TiS_2$, these two compounds also exhibit absorption in the UV region.

The reflectance of pristine $TiS_2$ supercell from Figure 4(b) or 5(c, d) & that of $TiSe_2$ and $TiTe_2$ from Table 7 shows that these compounds show moderate amount of reflectivity. For $TiS_2$, highest reflectance attained is 0.47 at 2.29 eV and that of $TiSe_2$ & $TiTe_2$ is 0.43 at 1.39 and 0.61 at 2.38 eV. The range of peaks for $TiS_2$ and $TiTe_2$ is in visible region & in near infrared region for $TiTe_2$. So, ~50-60 % reflection of light in visible and infrared region is observed for these compounds and hence it shows they reflect more than half of the incident light.

**Effect of Electric field** – By applying different magnitudes of $E_z$, there is a gradual increase in the absorption coefficient observed for all the compounds (Table 6). For $TiS_2$, $\alpha$ increases maximum to $5.7 \times 10^5$ cm$^{-1}$ at 11.32 eV energy with field strength of 10 V/nm. Similarly for $TiSe_2$ and $TiTe_2$, at the same field strength, $\alpha$ increases to $10.7 \times 10^5$ and $8.3 \times 10^5$ cm$^{-1}$ at 13.84 eV & 13.98 eV respectively. The range of energies with highest peaks observed for all the values of $E_z$ shows they all lay in the UV region only for all the compounds. However, when the energy value with the highest peak is considered, some blue-shifts and red-shifts are observed. For $TiS_2$, there is a red-shift of ~5.5 eV with $E_z$ values ranging from 2.5 – 7.5 V/nm. Again in $TiSe_2$, minor blue-shifts of 0.1-0.6 eV for all values of $E_z$ and in $TiTe_2$, red shifts of ~5 eV is present for $E_z$ range of 5-10 V/nm. The reflectivity is somewhat tuned by increasing $E_z$. It has gradually ascended to 0.59 and 0.57 at 2.22 and 1.19 eV for $TiS_2$ & $TiSe_2$ respectively and descended to 0.42 at 2.25 eV for $TiTe_2$ for field strength of 10 V/nm (Table 7). For $TiS_2$, the reflectance region is shifted to near IR from visible region at 2.5-7.5 V/nm causing a red-shift of ~0.7 eV. In $TiSe_2$ and $TiTe_2$ the region is same with red-shifts of ~0.1-0.2 eV and ~0.1 eV respectively.

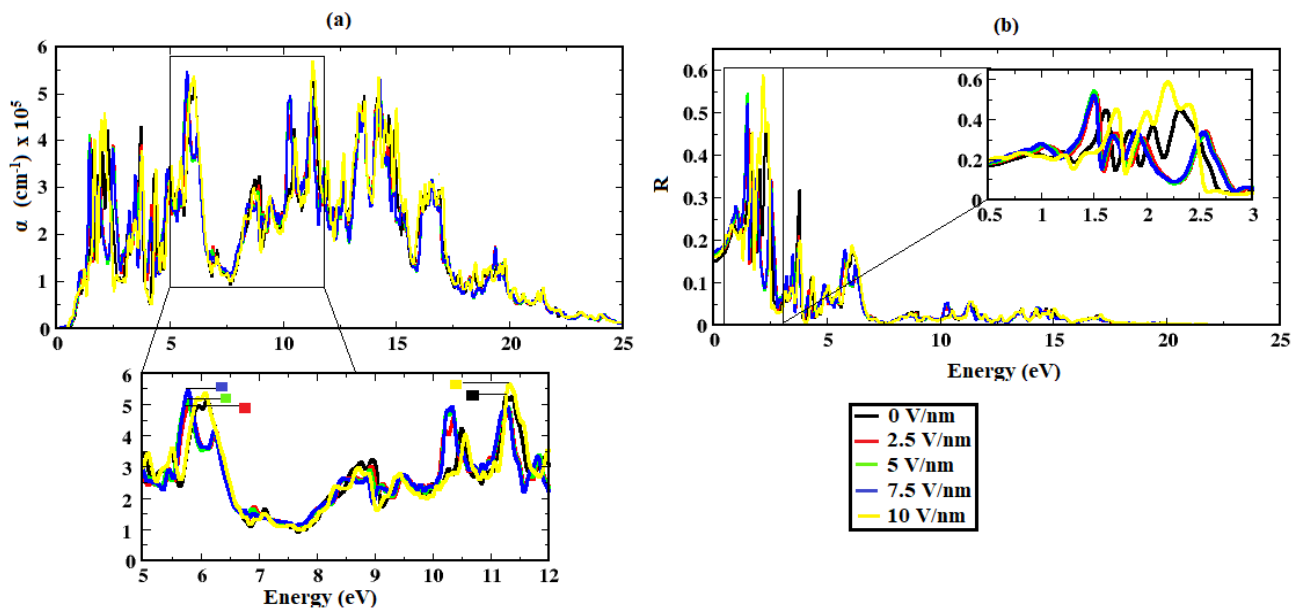

Figure 4 – (a) Absorption coefficient and (b) Reflectance of $TiS_2$ under the effect of electric field. The peak regions are shown in the inset. The peaks of each field strength is shown in respective colour in inset of fig 2(a)

| $E_z$ (V/nm) & Strain (%) | TiS$_2$ | | | TiSe$_2$ | | | TiTe$_2$ | | |
|---|---|---|---|---|---|---|---|---|---|
| | Energies with highest peaks (eV) | Energy at max. peak (eV) | Maximum α (cm$^{-1}$) x 10$^5$ | Energies with highest peaks (eV) | Energy at max. peak (eV) | Maximum α (cm$^{-1}$) x 10$^5$ | Energies with highest peaks (eV) | Energy at maximum peak (eV) | Maximum α (cm$^{-1}$) x 10$^5$ |
| Pristine supercell | 6.06, 11.37 | **11.37** | **5.0** | 10.34, 13.22 | **13.22** | 10.4 | 8.75, 13.82 | **13.82** | 6.5 |
| 2.5 V/nm | 5.81, 11.18 | **5.81** | 5.1 | 10.42, 13.31 | **13.31** | 10.4 | 8.69, 13.89 | **13.89** | 6.4 |
| 5 V/nm | 5.73, 11.23 | **5.73** | 5.2 | 10.44, 13.52 | **13.52** | 10.5 | 8.73, 13.91 | **8.73** | 6.6 |
| 7.5 V/nm | 5.75, 11.29 | **5.75** | 5.5 | 10.46, 13.68 | **13.68** | 10.6 | 8.77, 13.92 | **8.77** | 7.9 |
| 10 V/nm | 6.06, 11.35 | **11.35** | 5.7 | 10.49, 13.86 | **13.86** | 10.7 | 8.82, 13.95 | **8.82** | 8.3 |
| +2 % | 12.8-15.0 | **13.69** | 4.9 | 12.7-14.2 | **13.68** | 4.5 | 9.7-11.0 | **10.57** | 4.0 |
| +5 % | 12.5-13.5 | **13.05** | 4.8 | 13.3-13.8 | **13.74** | 4.1 | 8.5-10.8 | **9.59** | 4.5 |
| +10 % | 12.7-13.5 | **12.83** | 5.2 | 9.8-11.0 | **10.24** | 5.6 | 8.8-10.6 | **9.56** | 4.9 |
| -2 % | 13.0-15.0 | **14.21** | 4.9 | 13.0-14.5 | **13.19** | 4.9 | 10.4-12.0 | **11.04** | 4.7 |
| -5 % | 12.2-14.5 | **14.23** | 5.1 | 13.2-15.0 | **14.18** | 5.5 | 11.9-13.5 | **12.55** | 5.5 |
| -10 % | 12.0-14.6 | **14.41** | 5.5 | 7.5-9.3 | **8.71** | 9.6 | 11.5-13.3 | **12.96** | 5.8 |

Table -6 Energies with peak positions and absorption coefficient values under the effect of $E_z$ and strain for all the compounds. ('-' represents range of energies with highest peak & ',' represents particular energies of high peaks)

**Effect of strain –** Both the tensile & compressive strain tends to alter the absorption and there are no particular sharp peaks observed. The α value increases with increasing the amount of strain (Table 6). For TiS$_2$, α increases to 5.2 & 5.5 x 10$^5$ cm$^{-1}$ at 12.83 & 14.41 eV energy peaks for +10 % and -10 % strain respectively. For same amounts of tensile and compressive strain, again for TiSe$_2$ and TiTe$_2$, α increases to 5.6 & 9.6 x 10$^5$ cm$^{-1}$ at 10.24 & 8.71 eV and 4.9 & 5.8 x 10$^5$ cm$^{-1}$ at 9.56 & 12.96 eV respectively. The recorded energies are at UV region, as similar to pristine supercell. Again, for TiS$_2$, there is a blue-shift of ~1.5-2.3 eV, for TiSe$_2$, there is blue shift of ~0.5-1 eV and red shift of ~3-4.5 eV and for TiTe$_2$ there is only red shift of ~1-4 eV. The increase in absorption and the shifts is seen due to excitation of phonons by increasing both tensile & compressive applied strain.

The reflectance values from Table 7 shows that there are some sharp peaks for some values of applied strain, where there is shift in the energy. However, despite such jumps, it can be seen that the reflectance values are almost negligible in most of the cases. For TiS$_2$, R value ranges from 0.06-0.16 for both the directions of strain. For TiSe$_2$ and TiTe$_2$, it ranges from 0.06-0.26 & 0.08-0.27 respectively for both tensile & compressive strain. Also, there is blue-shift of ~0.5-2.5 eV and red-shift of ~1 eV for TiS$_2$. In case of TiSe$_2$, there is blue shift of ~1 eV and red-shift of ~1.2-1.3 eV and for TiTe$_2$ there is only red shift of ~1-2.3 eV. So, with the applied strain, it can be seen that reflectivity is drastically reduced.

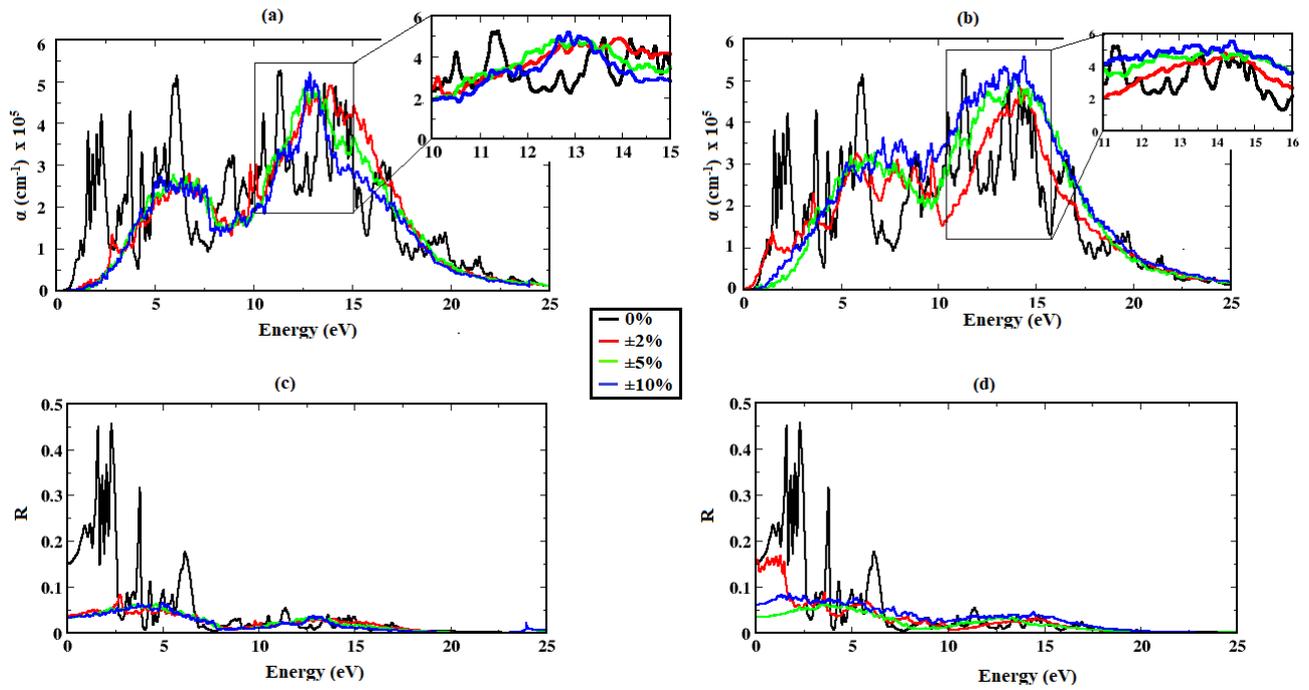

Figure – 5 Absorption coefficient and Reflectance of TiS$_2$ under (a, c) tensile and (b, d) compressive strain. The peak regions are shown in the inset.

| $E_z$ (V/nm) & Strain (%) | TiS$_2$ | | | TiSe$_2$ | | | TiTe$_2$ | | |
|---|---|---|---|---|---|---|---|---|---|
| | Energies with highest peaks (eV) | Energy at max. peak (eV) | Maximum R | Energies with highest peaks (eV) | Energy at max. peak (eV) | Maximum R | Energies with highest peaks (eV) | Energy at maximum peak (eV) | Maximum R |
| Pristine supercell | 1.63, 2.29 | 2.29 | **0.47** | 1.39, 1.71 | 1.39 | **0.43** | 2.38, 3.40 | 2.38 | **0.61** |
| 2.5 V/nm | 1.59, 2.62 | 1.59 | **0.52** | 1.30, 1.86 | 1.30 | **0.48** | 2.31, 3.30 | 2.31 | **0.55** |
| 5 V/nm | 1.57, 2.60 | 1.57 | **0.53** | 1.28, 1.83 | 1.28 | **0.52** | 2.28, 3.27 | 2.28 | **0.50** |
| 7.5 V/nm | 1.49, 2.59 | 1.49 | **0.51** | 1.23, 1.75 | 1.23 | **0.55** | 2.26, 3.17 | 2.26 | **0.45** |
| 10 V/nm | 2.22, 2.41 | 2.22 | **0.59** | 1.19, 1.71 | 1.19 | **0.57** | 2.25, 3.15 | 2.25 | **0.42** |
| +2 % | 2.5-2.8 | 2.78 | **0.08** | 0.06-0.19 | 0.15 | **0.12** | 1.1-1.5 | 1.40 | **0.08** |
| +5 % | 4.5-4.8 | 4.65 | **0.06** | 2.37-4.14 | 3.34 | **0.07** | 0.3-0.6 | 0.40 | **0.09** |
| +10 % | 4.7-5.1 | 5.01 | **0.06** | 2.48, 10.27 | 2.48 | **0.06** | 1.28, 1.48 | 1.28 | **0.10** |
| -2 % | 0.6-1.4 | 1.26 | **0.16** | 0.57-0.1 | 0.94 | **0.10** | 0.4-0.5 | 0.46 | **0.11** |
| -5 % | 3.3-3.8 | 3.43 | **0.06** | 0.02 | 0.02 | **0.24** | 0.06 | 0.06 | **0.23** |
| -10 % | 1.3-5.2 | 1.77 | **0.08** | 0.11, 8.80 | 0.11 | **0.26** | 0.1-0.2 | 0.11 | **0.27** |

Table -7 Energies with peak positions and reflectance values under the effect of $E_z$ and strain for all the compounds. ('-' represents range of energies with highest peak & ',' represents particular energies of high peaks)

### (c) Optical conductivity and refractive index

The maximum optical conductivity for all the pristine supercell structures is nearly equal with TiTe$_2$ having the highest value of $\sigma = 2.81 \times 10^5$ $\Omega m^{-1}$. Also, the energies with maximum peaks lie in the near IR region for TiS$_2$ & TiSe$_2$ but some peaks are also present in the visible region for TiTe$_2$ (Table 8). However, some peaks of comparatively lesser σ values were seen in the visible region also in case of TiS$_2$ and TiSe$_2$. With increase in photon energy, conductivity seems to get vanished. The range of peaks attained for conductivity is aligned to the absorption spectra.

The static refractive index of pristine supercell of all the compounds is almost same. But the maximum value of $n$ attained is 2.26, 3.41 and 3.76 for $TiS_2$, $TiSe_2$ & $TiTe_2$ respectively, with the energies at the highest peak being almost equal for them. The refractive index is > 1 due to the fact that as photons enter a material they are slowed down by the interaction with electrons and further, the more photons are slowed down travelling via the material, more is the refractive index. The peaks with high energies are mainly seen in the near IR region only for all the compounds (Table 9).

**Effect of electric field –** With the electric field, for $TiS_2$ and $TiSe_2$ the $\sigma$ value increases and for $TiTe_2$, it is decreased. In case of $TiS_2$, $\sigma$ increases maximum to 2.74 x $10^5$ $\Omega m^{-1}$ at 1.44 eV for 5 V/nm (Figure 6(a))and to 3.10 x $10^5$ $\Omega m^{-1}$ at 1.13 eV for 10 V/nm in case of $TiSe_2$ and decreases to 2.46 x $10^5$ $\Omega m^{-1}$ at 1.51 eV for 7.5 V/nm in case of $TiTe_2$. Also, there is very minor energy shifts for $TiS_2$ and $TiSe_2$. For $TiTe_2$, there is no shift in energy peaks is observed until $E_z$ = 10 V/nm, where a blue shift of ~7 eV is seen. However, having a look at the energies with highest peaks on applying the external field, it is observed that high peaks of $\sigma$ values are present in both near IR & visible as well as UV regions.

Electric field seems to modulate the refractive index for all the compounds. The static value of $n$ is gradually increased by applying $E_z$ in all the cases, with increased value 2.40, 2.48 & 2.51 for $TiS_2$, $TiSe_2$ & $TiTe_2$ respectively at 10 V/nm. Whereas, with increasing strength of $E_z$ the maximum value of $n$ increases abruptly for $TiS_2$ and smoothly for $TiSe_2$. On the other hand, it decreases smoothly for $TiTe_2$ (Table 6). The region of highest energy peaks are also the same, which lies in the near IR regime. This also shows that there is no major energy shifts. The increase in static $n$ is due to the increase in electron density caused by $E_z$ and any factor increasing the electron density also increase the refractive index. With increase in photon energy, the trend is altered because of electronic excitations.

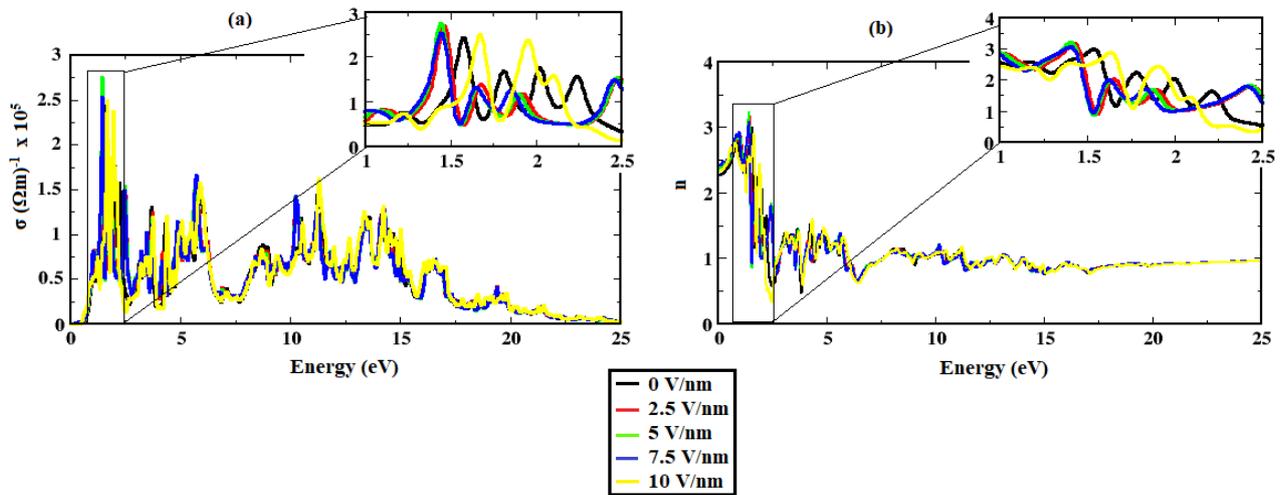

Figure – 6 (a) Optical conductivity and (b) Refractive index of $TiS_2$ under the effect of electric field. The peak regions are shown in the inset.

| | $TiS_2$ | | | $TiSe_2$ | | | $TiTe_2$ | | |
|---|---|---|---|---|---|---|---|---|---|
| $E_z$ (V/nm) & Strain (%) | Energies with highest peaks (eV) | Energy at max. peak (eV) | Maximum $\sigma$ ($\Omega m^{-1}$) x $10^5$ | Energies with highest peaks (eV) | Energy at max. peak (eV) | Maximum $\sigma$ ($\Omega m^{-1}$) x $10^5$ | Energies with highest peaks (eV) | Energy at maximum peak (eV) | Maximum $\sigma$ ($\Omega m^{-1}$) x $10^5$ |
| Pristine supercell | 1.57, 1.65 | 1.57 | **2.44** | 1.31, 1.71 | 1.31 | **2.28** | 1.51, 2.15 | 1.51 | **2.81** |
| 2.5 V/nm | 1.46, 2.46 | 1.46 | **2.69** | 1.24, 1.67 | 1.24 | **2.71** | 1.51, 2.05 | 1.51 | **2.70** |
| 5 V/nm | 1.44, 2.46 | 1.44 | **2.74** | 1.20, 1.67 | 1.20 | **2.89** | 1.51, 8.71 | 1.51 | **2.61** |
| 7.5 V/nm | 1.43, 2.45 | 1.43 | **2.52** | 1.16, 1.66 | 1.16 | **3.02** | 1.51, 8.74 | 1.51 | **2.46** |
| 10 V/nm | 1.68, 1.95 | 1.68 | **2.50** | 1.13, 1.69 | 1.13 | **3.10** | 1.39, 8.77 | 8.77 | **2.72** |
| +2 % | 11.6-14.0 | 12.99 | **1.34** | 12.5-14.5 | 13.71 | **1.19** | 9.0-10.6 | 9.20 | **1.44** |
| +5 % | 11.6-13.5 | 12.52 | **1.39** | 11.5-13.7 | 11.95 | **1.08** | 9.40, 11.38 | 9.40 | **1.13** |
| +10 % | 11.5-13.2 | 12.68 | **1.44** | 10.24, 12.18 | 10.24 | **1.75** | 9.47, 10.56 | 9.47 | **1.28** |
| -2 % | 12.7-15.0 | 13.52 | **1.12** | 11.7-14.2 | 13.14 | **1.28** | 8.5-11.3 | 11.07 | **1.24** |
| -5 % | 11.2-15.0 | 12.52 | **1.34** | 13.4-14.5 | 13.99 | **1.45** | 7.17, 11.95 | 7.17 | **1.46** |
| -10 % | 10.5-15.0 | 13.17 | **1.36** | 8.05, 8.70 | 8.70 | **1.51** | 8.95, 12.10 | 8.95 | **1.98** |

Table -8 Energies with peak positions and optical conductivity values under the effect of $E_z$ and strain for all the compounds. ('-' represents range of energies with highest peak & ',' represents particular energies of high peaks)

**Effect of strain –** The application of both tensile and compressive strain shows that the σ value is reduced by stretching the lattice in case of all the compounds. Some large amount of blue-shifts is also seen (Table 8). For the compounds, σ value is almost same for strain applied along both directions and for $TiS_2$, it reduces to 1.34 & 1.12 x $10^5$ $\Omega m^{-1}$ for +2% & -2% respectively. For $TiSe_2$, σ is decreased to 1.08 & 1.28 x $10^5$ $\Omega m^{-1}$ for +5% & -2% respectively and for $TiTe_2$, for same amount of strain, σ is reduced to 1.13 & 1.24 $\Omega m^{-1}$. For both the direction of strain, there is a blue-shift of ~12 eV, ~7.5-12.5 eV and ~6-9.5 eV in case of $TiS_2$, $TiSe_2$ and $TiTe_2$ respectively. Similar to what observed with applied $E_z$, peaks is observed near IR, visible as well as UV regions.

For refractive index, the value of static *n* as well as maximum value of *n* is decreased in case of tensile strain (Table 9). For $TiS_2$, there is also some blue-shifts in energy (~1 eV) is seen for tensile strain and red-shifts of 0.75-1 eV is seen in case of $TiTe_2$. For $TiSe_2$, blue-shift of ~1 eV is observed for +5% and +10% strain & a red-shift of ~1eV for +2% strain. When compressive strain is applied, for $TiS_2$, both the static and maximum value of *n* is reduced but less as compared to tensile strain. A similar observation is recorded for maximum *n* for the other two compounds, but for both $TiSe_2$ and $TiTe_2$, static *n* increases for -5% & -10% strain. Also, applying compressive strain, we see similar values of static and maximum *n*, which also causes energy shifts to mid-IR region.

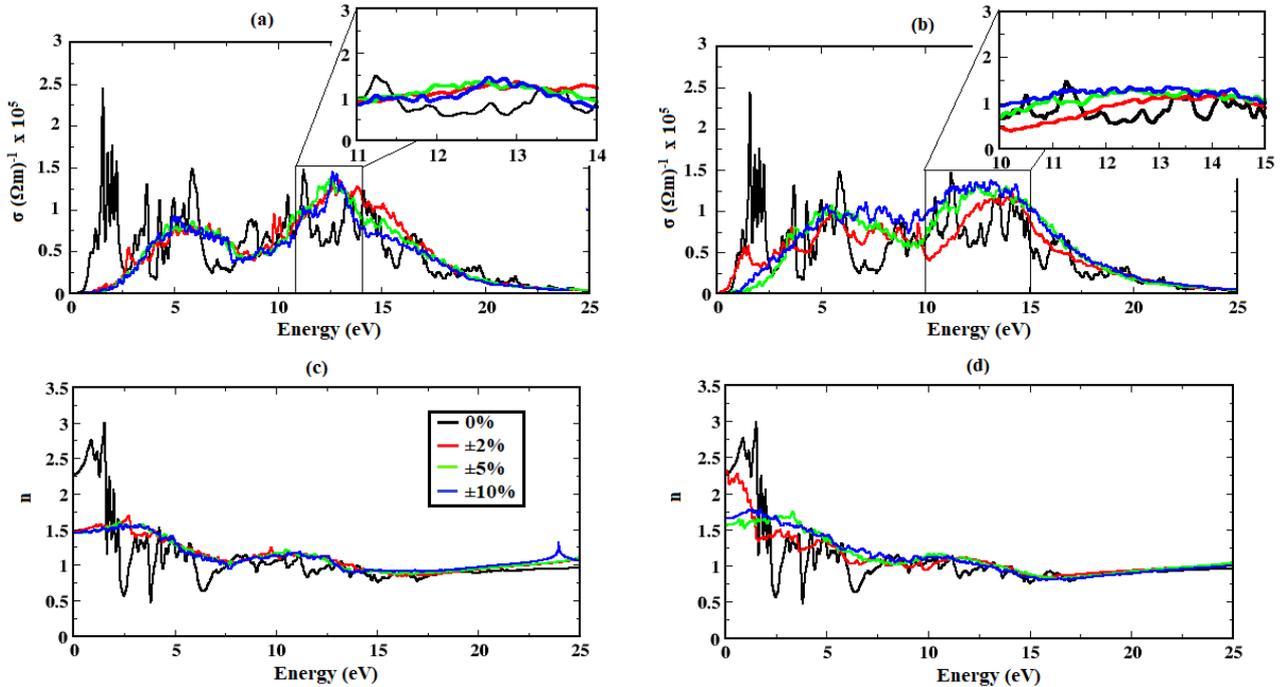

Figure – 7 Optical conductivity and Refractive index of $TiS_2$ under (a, c) tensile and (b, d) compressive strain. The peak regions are shown in the inset.

| $E_z$ (V/nm) & Strain (%) | TiS$_2$ | | | TiSe$_2$ | | | TiTe$_2$ | | |
|---|---|---|---|---|---|---|---|---|---|
| | Energy at max. peak (eV) | Maximum $n$ | Static refract. Index $n$ | Energy at max. peak (eV) | Maximum $n$ | Static refract. Index $n$ | Energy at max. peak (eV) | Maximum $n$ | Static refract. Index $n$ |
| Pristine supercell | 1.53 | 2.99 | 2.26 | 1.24 | 3.41 | 2.27 | 1.48 | 3.76 | 2.30 |
| 2.5 V/nm | 1.42 | 3.20 | 2.32 | 1.17 | 3.81 | 2.35 | 1.35 | 3.69 | 2.38 |
| 5 V/nm | 1.40 | 3.22 | 2.34 | 1.13 | 3.99 | 2.39 | 1.31 | 3.67 | 2.42 |
| 7.5 V/nm | 1.39 | 3.06 | 2.37 | 1.10 | 4.17 | 2.45 | 1.26 | 3.64 | 2.46 |
| 10 V/nm | 1.63 | 3.01 | 2.40 | 1.07 | 4.28 | 2.48 | 1.24 | 3.59 | 2.51 |
| +2 % | 2.72 | 1.68 | 1.47 | 0.13 | 2.04 | 2.01 | 0.86 | 1.77 | 1.71 |
| +5 % | 2.27 | 1.59 | 1.44 | 1.96 | 1.68 | 1.51 | 0.35 | 1.87 | 1.82 |
| +10 % | 2.32 | 1.56 | 1.43 | 2.15 | 1.64 | 1.45 | 1.27 | 1.84 | 1.64 |
| -2 % | 0.02 | 2.25 | 2.25 | 0.56 | 1.81 | 1.74 | 0.36 | 1.99 | 1.97 |
| -5 % | 3.33 | 1.74 | 1.55 | 0.00 | 2.98 | 2.98 | 0.00 | 2.91 | 2.91 |
| -10 % | 1.74 | 1.78 | 1.64 | 0.00 | 3.15 | 3.15 | 0.00 | 3.22 | 3.22 |

Table -9 Values of static and maximum refractive index including energies with peak positions under the effect of $E_z$ and strain for all the compounds.

The increase in some of the optical properties observed due to $E_z$ is mainly because of increased electron density movement causing interband transitions starting from lower photon energy to higher photon energy. All the interband transitions are essentially due to the p orbital of the X (X = S, Se & Te) atoms moving to the d orbital of the Ti atoms. When both tensile and compressive is applied, it is observed that although there is a shift in the energy with the maximum value of the properties, there are very less sharp peaks obtained and thus the shift in energy has not occurred swiftly, rather a gradual rise in the peaks is seen. This shows hindrance in the interband transitions which occurs because electron correlation plays a major role for the strained structures and the electrons are localized around the atoms, thus becoming less drifting. So, it can be seen a reduction in the obtained values of properties with strain. However, for compressive strain, due to reduction in Ti-X distance (Table 2), there is redistribution of electrons which may enable orbital excitations. Hence, some sharp peaks and increase in values of properties can be seen for compressive strain.

## Conclusion

In summary, the magnetic and optical properties of TiX$_2$ (X= S, Se and Te) compounds have been calculated using ab-initio calculations including the effect of externally applied electric field and biaxial strain. A 3x3 supercell of the compounds has been taken to study the properties. Optimization of the structures including magnetic and non-magnetic states has been performed and it is found that the magnetic state (ferromagnetic nature) is more stable. So using spin polarized calculations, the properties of the pristine supercells is investigated. It is observed that a considerable amount of magnetic moment is attained for the supercell structures. Also, the compounds show good response to the deduced optical properties, which is calculated using polarized light along z-direction. Further, electric field (along z-direction) and biaxial tensile and compressive strain (along x and y-direction) is employed on the structures. With electric field, it is seen that there is increase in the magnetic moment as well as the calculated optical properties, which is observed due to the increase in orbital transitions of electrons in the compounds. On the other hand, for applied tensile strain, the calculated values of both the properties are seen to reduce. For compressive strain, magnetic moment increases to a good extent and some of the optical properties (static $n$ & static $\varepsilon$ and $\alpha$ in some of the cases) are also seen to increase with increasing the strain. But for most of the optical properties are observed to decrease under applied compressive strain. Some major shifts in energies with highest peaks of these properties are also seen when strain is applied, where the peaks shift from one region of the electromagnetic spectrum to other.

Hence, applying these external factors, we can attain quite good amount of magnetic moment and even be able to tune it, which shows these compounds can be widely used in the field of spintronics. Also, with the factors into effect, we can obtain modulating optical properties and especially the attained high refractive index and dielectric constant can find applications in optoelectronic devices.